\documentclass[%
 aip,
 amsmath,amssymb,
 reprint,%
]{revtex4-1}

\usepackage[version=3]{mhchem} 
\usepackage{amsmath}
\usepackage{amsfonts}
\usepackage{amssymb}

\usepackage[utf8]{inputenc}
\usepackage[T1]{fontenc}
\usepackage{mathptmx}
\usepackage{etoolbox}

\usepackage[utf8]{inputenc}
\usepackage[T1]{fontenc}

\usepackage{amsmath}
\usepackage{amsthm}
\usepackage{amsfonts}
\usepackage{amssymb}
\usepackage{mathtools}
\usepackage{chemformula}
\usepackage{multirow}

\theoremstyle{plain}

\theoremstyle{definition}

\theoremstyle{remark}

\theoremstyle{example}

\usepackage{titlesec}

\begin{document}

\newcommand{\nyuchem}{\affiliation{\mbox{Department of Chemistry, New York University, New York, N.Y. 10003}}}
\newcommand{\simons}{\affiliation{\mbox{Simons Center for Computational Physical Chemistry, New York, N.Y. 10003}}}
\newcommand{\cds}{\affiliation{\mbox{Center for Data Science, New York University, New York, N.Y. 10004}}}
\newcommand{\sh}{\affiliation{Shanghai Frontiers Science Center of Artificial Intelligence and Deep Learning and NYU-ECNU Center for Computational Chemistry, NYU Shanghai, Shanghai 200062, P.R.China}}

\title{
    Molecular Dynamics and Machine Learning Unlock Possibilities in Beauty Design---A Perspective
}

\author{Yuzhi Xu}
\simons
\nyuchem
\sh

\author{Haowei Ni}
\simons

\author{Fanyu Zhao}
\sh

\author{Qinhui Gao}
\affiliation{
\mbox{Department of Digital Humanities, King’s College London, Strand, London WC2R 2LS, U.K.
}}

\author{Ziqing Zhao}
\author{Chia-Hua Chang}
\author{Yanran Huo}
\simons

\author{Shiyu Hu}
\sh

\author{Yike Zhang}
\affiliation{\mbox{Division of Cardiology,
The First Affiliated Hospital of Nanjing Medical University, Nanjing 210029, P.R.China}}

\author{Radu Grovu}
\affiliation{\mbox{Internal Medicine Department, Rhode Island Hospital, Brown Univerisity Health, Providence, R.I., 02903
}}

\author{Min He}
\affiliation{
Xbiome, Inc., Cambridge, Mass. 01451
}

\author{John Z. H. Zhang}
\sh
\nyuchem

\author{Yuanqing Wang}
\email[Correspondence email address: ]{wangyq@wangyq.net}
\simons
\nyuchem
\cds

\begin{abstract}
Computational molecular design---the endeavor to design molecules, with various missions, aided by machine learning and molecular dynamics approaches, has been widely applied to create valuable new molecular entities, from small molecule therapeutics to protein biologics.
In the small data regime, physics-based approaches model the interaction between the molecule being designed and proteins of key physiological functions, providing structural insights into the mechanism.
When abundant data has been collected, a quantitative structure-activity relationship (QSAR) can be more directly constructed from experimental data, from which machine learning can distill key insights to guide the design of the next round of experiment design.
Machine learning methodologies can also facilitate physical modeling, from improving the accuracy of force fields and extending them to unseen chemical spaces, to more directly enhancing the sampling on the conformational spaces.
We argue that these techniques are mature enough to be applied to not just extend the \textit{longevity} of life, but the \textit{beauty} it manifests.
In this perspective, we review the current frontiers in the research \& development of skin care products, as well as the statistical and physical toolbox applicable to addressing the challenges in this industry.
Feasible interdisciplinary research projects are proposed to harness the power of machine learning tools to design innovative, effective, and inexpensive skin care products.

\end{abstract}

\maketitle




\section{Introduction: Beauty design and computer-aided molecular discovery.}
The human race has been using cosmetics and skin-care products for more than 8000 years~\cite{britannicaStartWearing}---the history of which arguably predates that of medical practice.
Through these 8 millenniums, while medical and medicinal sciences have been transformed into quantitative, mechanistic, rigorous, and digitalized disciplines, the design of cosmetics and skin-care products, referred to as \textit{beauty design} henceforth, remains largely empirical, experimental, and trial-and-error-based.

The \textit{in silico} approaches to design new molecule entities, namely therapeutics, using computer simulation and modeling are collectively termed \textit{computer-aided} molecular discovery.
Roughly, they can be characterize into two categories, \textit{physics-inspired} and \textit{data-driven}, which correspond to \textit{ligand-based} and \textit{structure-based} drug discovery in medicinal chemistry, respectively.
The former models the \textit{conformational} landscape of a biomolecular system $\mathcal{G}$, which adopts a Boltzmann-type distribution:
\begin{equation}
\label{eq:boltzmann}
p(\mathbf{x}|\mathcal{G}) \propto \exp(-u(\mathbf{x}; \mathcal{G})),
\end{equation}
based on which ensemble properties of the system can be estimated using the samples drawn from this distribution $\mathbf{x}_i$:
\begin{equation}
\label{eq:md}
<\mathcal{O}(\mathbf{x})>
= \int \mathrm{d} \mathbf{x} p(\mathbf{x}|\mathcal{G}) \mathcal{O}(\mathbf{x})
\approx \sum\limits_{\mathbf{x}_i} \mathcal{O}(\mathbf{x}_i).
\end{equation}
And molecular dynamics (MD) simulations can be employed to generate such samples.
If one only wishes to draw insights from the \textit{minima} of Equation~\ref{eq:boltzmann}, they can further simplify Equation~\ref{eq:md} to be:
\begin{equation}
\hat{\mathcal{O}}(\mathbf{x}) \approx \mathcal{O}(\operatorname{argmax}_i p(\mathbf{x}_i|\mathcal{G})),
\end{equation}
where the optima are found by structural methods such as docking~\cite{meng2011molecular} or protein folded pose prediction~\cite{jumper2021highly, abramson2024accurate}, although this can be a very crude approximation neglecting the dynamic nature of biomolecular systems.

On the other hand, \textit{data-driven} methods seek insights directly from data, most typically the molecule-measurement pairs.
Quantitative structure-activity relationship (QSAR) models are constructed to model the mapping $f: \mathcal{G} \rightarrow \mathbb{R}$ from the molecular topological graphs $\mathcal{G}$ to scalar physical, chemical, or physiological properties.
This is where machine learning has become an indispensable workhorse for designing new molecular entities~\cite{retchin2024druggym, dara2022machine, wang2023scientific}.
In principle, this school of molecular modeling characterizes the \textit{posterior predictive distribution}
\begin{equation}
\label{eq:posterior-predictive}
p(\hat{y}_\mathcal{G}|\mathcal{G}, \mathcal{M}) = \int \mathrm{d} \Theta p(\hat{y}_\mathcal{G} | \mathcal{G}, \Theta, \mathcal{M}) p(\Theta | \mathcal{M}, \mathcal{D}),
\end{equation}
where the property of a molecule $\mathcal{G}$ given a model $\mathcal{M}$ can be written as that dependent upon the parameters to the model $\Theta$, marginalized across the parameter distribution conditioned on the existing data $\mathcal{D} = \{ \mathcal{G}_i, y_i \}$.
This principled Bayesian approach is usually replaced with a single point maximum-likelihood estimate (MLE) on the parameter space.

While traditional fingerprint-based~\cite{wigh2022review} methods are still used, small molecules are now more ubiquitously modeled as graphs, and processed through a graph neural network (GNN)~\cite{DBLP:journals/corr/KipfW16, xu2018powerful, gilmer2017neural, hamilton2017inductive, battaglia2018relational, wang2024nonconvolutionalgraphneuralnetworks} to form useful representations for downstream prediction tasks.
Once the backbone of this mapping is constructed, it can also be used to quantify the uncertainty of its predictions in a Bayesian setting~\cite{wang2021stochasticaggregationgraphneural, wangbayesian}, for example with variational inference~\cite{Blei_2017}, or in an active learning setting to guide the design of the next round of experiments~\cite{kristiadi2024soberlookllmsmaterial}.

Additionally, machine learning techniques are also applied in a \textit{structure-based} molecular discovery setting, where the accuracy or efficiency of traditional physical models are revolutionized by machine learning methodologies.
On the one hand, machine learning provides new functional forms~\cite{Smith_2017, abramson2024accurate, schütt2021equivariant, thölke2022torchmdnet, Unke_2019, eastman2023openmm, wang2023spatial} and parametrization methods~\cite{wang2022end, takaba2024machine, wang2024open, wang2024espalomacharge, wang2024designspacemolecularmechanics} for force fields to be used in molecular dynamics (MD) simulations.
On the other hand, it has also been applied to directly predict the structure or conformational distributions~\cite{noé2019boltzmanngeneratorssampling}.
When it comes to proteins, in particular, the determination of the folded conformation thereof has been dramatically accelerated using machine learning-based models~\cite{jumper2021highly, abramson2024accurate, ahdritz2024openfold}, and with the predicted structures one can readily \textit{design}~\cite{kuhlman2003design} new proteins to fulfill desired properties or functions.

In this perspective, we first review the overarching goals and specific challenges in beauty design.
Next, we identify the physical (MD-based) and machine learning toolboxes useful in modeling and optimizing the properties of interest in beauty design.
Finally, we also envision beauty design areas where MD simulation and ML modeling can be of immediate utility, and propose feasible projects.

\begin{figure*}
    \centering
    \includegraphics[width=0.9\linewidth]{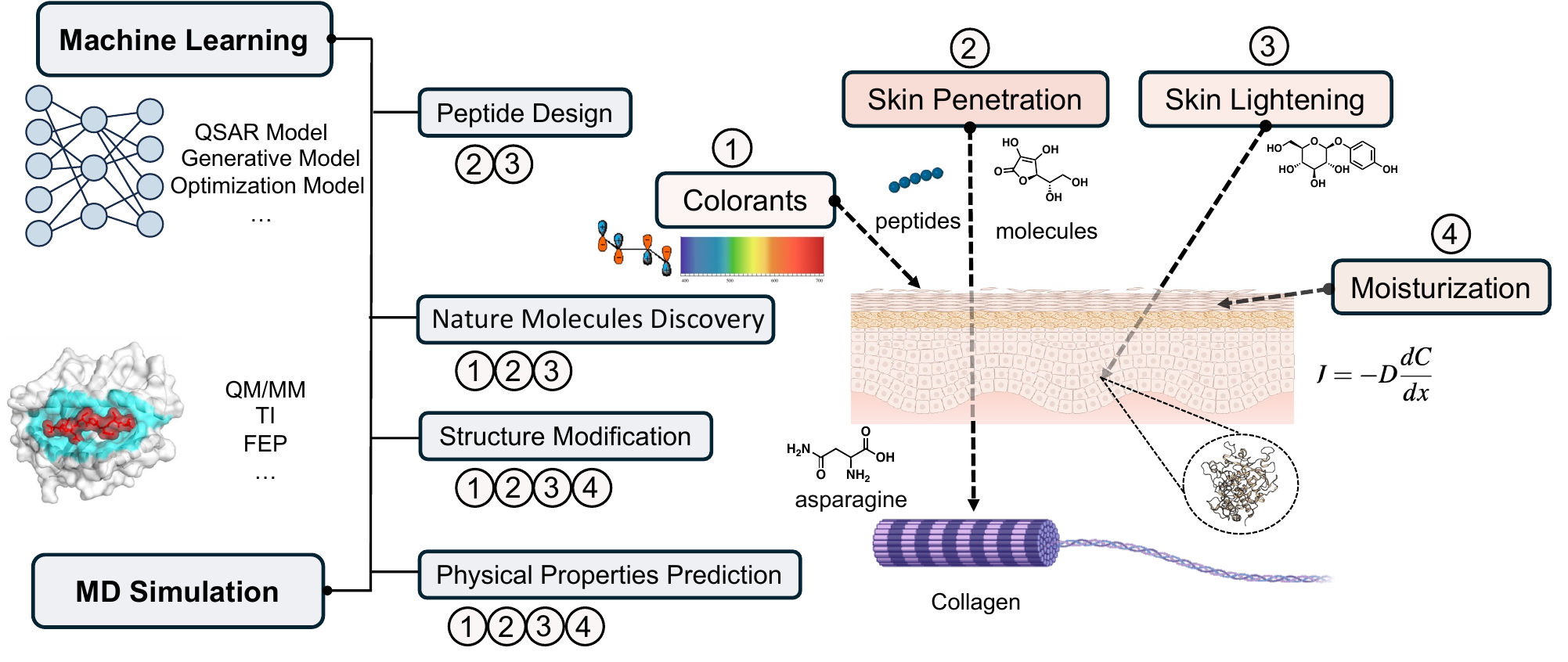}
    \caption{\textbf{Semantic overview.}}
    \label{fig:abstract}
\end{figure*}

\section{Objectives in skincare and cosmetics}
\label{sec:axes}
In this section, we review the primary goals in beauty design, the general taxonomy, and the physical laws to be considered in each axis.

\subsection{Coloration.}
Cosmetic colorants can be classified into two major categories based on their chemical composition: inorganic and organic colorants~\cite{hefford2011colourants}. 
The global annual production of inorganic colorants is approximately 10 million tons, accounting for 80\% of the total colorant production~\cite{lewis1998colorants}. 
Key inorganic colorants include titanium dioxide, iron oxides, carbon black, zinc oxide, ultramarine pigments, and mica-based pearlescent pigments, most of which are derived from natural minerals~\cite{giancola2015decorative,neytal2024advances}---these pigments are widely used in facial, eye, lip, and nail products. 
Due to their stable crystalline structure, inorganic pigments exhibit high stability, excellent opacity from their high refractive index, and natural coloration~\cite{lam2021factors}. 
Nevertheless, they also have certain limitations, such as a narrow range of light absorption, making it difficult to cover a broader visible spectrum, high surface energy leading to aggregation, and large particle size affecting skin permeability~\cite{gurses2016colorants}. 

As to organic colorants, they can be further divided into natural and synthetic colorants~\cite{zollinger2003color}.
Natural colorants are typically derived from plant and insect pigments, such as carmine, carotenoids, anthraquinones, betalains, and flavonoids. 
These colorants are popular among consumers due to their natural origin~\cite{sigurdson2017natural}, although they also face challenges such as color instability and susceptibility to environmental degradation. 
In contrast, synthetic colorants generally offer excellent color stability, a wide range of spectral options, and superior light and heat resistance~\cite{lanzendorfer2021colorants}. 
This is mainly due to their molecular structures, which often contain a large number of conjugated double bonds (e.g., azo and bonds, triphenylmethane structures). 
These conjugated systems extend electron delocalization, enabling absorption over a broader wavelength range of visible light and producing vivid colors~\cite{lewis1998colorants}. 
The highly conjugated structures not only enhance molecular stability and reduce structural breakdown under light or heat exposure but also enable optimal electronic transitions, effectively absorbing specific wavelengths of light to improve color vibrancy and stability~\cite{kiernan2001classification}. 
Synthetic colorants encompass a wide variety of types, including azo dyes, triphenylmethane dyes, flavonoid dyes, and anthraquinone dyes. 
Through optimized molecular design, these different types of dyes ensure color performance and stability across various application environments. 
Moreover, the consistency achieved through batch synthesis results in highly uniform molecular structures and purity, further enhancing the clarity and stability of the colors.
However, due to environmental and health concerns, the design of synthetic colorants requires more stringent molecular design criteria to minimize potential risks~\cite{renita2023progress}.

\subsection{Moisturization.}
The moisturizing efficacy of skincare products is typically assessed through Transepidermal Water Loss (TEWL), which measures the rate at which water diffuses from the stratum corneum into the atmosphere~\cite{rogiers2001eemco}, where an evaporimeter is used to quantify the amount of water lost per square meter of skin surface per hour, providing insights into the skin’s water kinetics. 
Since TEWL primarily reflects the process of water diffusing from the deeper layers of the skin through the epidermis to the external environment, this process can be approximately described by Fick’s First Law of Diffusion~\cite{rodrigues1999transepidermal}:
\begin{equation}
J = -D \frac{\mathrm{d}C}{\mathrm{d}x}
\end{equation}
where $J$ is the water flux 
$D$ is the diffusion coefficient of water in the stratum corneum 
and $\mathrm{d}C / \mathrm{d}x$ is the concentration gradient of water 
.

Based upon this relationship, different types of moisturizing products regulate TEWL by adjusting the diffusion coefficient \( D \) or the concentration gradient $\mathrm{d}C / \mathrm{d}x$, thus achieving a moisturizing effect. 
Typically, moisturizing ingredients in skincare products fall into one or more of the following three categories: humectants, occlusives, and emollients~\cite{neytal2024advances}. 
Humectants are generally compounds containing multiple hydroxyl groups (–OH), such as glycerin, hyaluronic acid, \(\alpha\) Hydroxy Acids, and propylene glycol. These molecules form stable hydrogen-bond networks with water, increasing the water content in the stratum corneum and lowering the chemical potential of water. 
By increasing the water concentration \( C \) within the stratum corneum, humectants reduce the concentration gradient $\mathrm{d}C / \mathrm{d}x$. 
According to Fick's first law, when humectants increase the water concentration in the stratum corneum, the concentration gradient $\mathrm{d}C / \mathrm{d}x$ decreases, leading to a reduction in the water flux \( J \) and consequently reducing the rate of water evaporation. 
Occlusives, such as petroleum jelly, mineral oil, and silicone oils, are typically hydrophobic compounds that form a physical barrier on the skin surface~\cite{de2020functional}. 
This barrier prevents water evaporation and thereby reduces the rate of water flux \( J \) from the skin, effectively minimizing transepidermal water loss. 
Emollients work by softening and smoothing the skin, filling the gaps between skin cells, many of which possess both occlusive and emollient properties.

\subsection{Skin Lightening}
To achieve skin lightening typically requires intervention in the processes of melanin production, transport, and metabolism in the skin~\cite{smit2009hunt}. 
Inhibiting melanin production is primarily achieved by targeting tyrosinase, a copper-containing metalloenzyme with two copper ions (CuA and CuB) in its active site, which are responsible for substrate binding and catalysis~\cite{lai2018structure,chang2009updated}. 
There are several strategies for inhibiting tyrosinase activity. 
The most direct approach involves inhibiting its active site. 
For example, arbutin effectively reduces tyrosinase activity through competitive inhibition, thereby decreasing melanin production~\cite{boo2021arbutin}. 
Another approach is to chelate the copper ions (CuA and CuB) at the active site, preventing the enzyme from functioning properly.
Kojic acid, for instance, can form chelates with the copper ions in tyrosinase, significantly inhibiting its catalytic function~\cite{chang2009updated}. 
Additionally, retinoic acid can suppress TYR gene promoter activity, reducing tyrosinase transcription and consequently lowering melanin synthesis at the source~\cite{zaidi2019natural}. 
Tyrosinase activity is also influenced by glycosylation modifications, and inhibitors like mannose can reduce tyrosinase glycosylation, thereby decreasing its enzymatic activity~\cite{petrescu1997inhibition}. 
Moreover, linoleic acid and $\beta$-sitosterol promote tyrosinase ubiquitination and proteasomal degradation, reducing the amount of active enzyme in the cell, and further inhibiting melanin production~\cite{logesh2023natural,shi2022ubiquitin}.

Blocking melanin transport is also an essential strategy for achieving skin whitening. By interfering with the transfer of melanosomes from melanocytes to surrounding keratinocytes, the distribution of melanin in the epidermis can be reduced. Niacinamide, for example, can inhibit protease-activated receptor-2 (PAR-2) on keratinocytes, thereby reducing melanosome transfer~\cite{smit2009hunt}. 
Similarly, licorice extract interferes with the melanosome transport process by inhibiting signaling pathways such as PI3K/Akt and MAPK, further reducing melanin deposition in the epidermis~\cite{wang2022licorice}.

\subsection{Skin Penetration.}
An important function in cosmetics is to activate collagen production and enhance antioxidant effects at the deeper layers of the skin~\cite{kusumawati2013natural,costa2017delivery}. The skin serves as a protective barrier against external environmental factors. 
Therefore, for cosmetic molecules to be effective, they must penetrate this barrier to reach the underlying layers. Common skin-penetrating ingredients in cosmetic products can be broadly categorized into small molecules, peptides and liposomes. 

In particular, small molecules in cosmetic products, such as Vitamin C, Vitamin A, Vitamin E, salicylic acid, and niacinamide, play crucial roles in processes such as amino acid hydroxylation in collagen, activation of signaling pathways, and antioxidation~\cite{murad1981regulation,mohiuddin2019skin,talakoub2016antiaging,sajna2015white,traber2007vitamin,dinarello2010anti}. 
Due to their molecular weights ranging from 122 to 430, these small molecules can penetrate the deeper layers of the skin through aqueous channels or the lipid bilayer. Despite their favorable permeability, the rate of penetration is influenced by several factors, including formulation, concentration, temperature, and pH. For example, Vitamin C requires a water-based formulation with a pH lower than its pKa (pKa 4.2) to reduce the charge density on the molecule~\cite{stamford2012stability}. Similarly, the skin permeability of salicylates exhibits a parabolic relationship with the partition coefficient, with optimal permeability typically occurring at a log P value of around 2.5. 
At this pH, salicylates have sufficient solubility in the lipid layer to facilitate diffusion through the stratum corneum while maintaining enough hydrophilicity to penetrate the viable epidermal tissue~\cite{benson2005transdermal}. 
To enhance the skin penetration of these small molecules while preserving their efficacy, various optimization strategies can be employed. These include molecular group modification, chemical potential optimization, eutectic systems, and coordination compounds.

In addition to their direct effects on the skin, small molecules also play a role in facilitating the penetration of peptides and proteins into the body. 
Peptides and larger protein molecules are particularly effective in wrinkle reduction, skin vitality maintenance, and collagen repair, as well as in sensitive factor attachment protein receptor (SNARE) regulation~\cite{rostkowska2023dermatological,lim2018enhanced}. 
However, due to their larger molecular weight and hydrophilic nature, these molecules face significant challenges in penetrating the hydrophobic lipid matrix of the stratum corneum~\cite{mortazavi2022skin}. 
To address these permeability issues, various strategies are being developed, such as the use of small-molecule chemical penetration enhancers (CPEs)~\cite{ahad2009chemical}.
These include azone derivatives, fatty acids, alcohols, esters, sulfoxides, pyrrolidones, glycols, surfactants, and terpenes~\cite{kumar2016proteins,karande2005design}. These chemicals can enhance the permeability of peptides and proteins by increasing the diffusion rate within the skin, inducing lipid fluidization, optimizing thermodynamic activity, influencing partition coefficients, and enhancing the release of drugs from formulations into the upper layers of the skin. However, while CPEs significantly improve permeability, they may also cause skin irritation and inflammation, and so far, no ideal chemical penetration enhancer has been identified~\cite{zeng2022molecular}.

In the way of searching for more effective, non-invasive, non-toxic, and non-irritating skin penetration enhancers, short-sequence peptides have been identified as safer alternatives to small molecule CPEs~\cite{kumar2015peptides,hmingthansanga2022improved}. 
By designing and synthesizing specific short-chain peptides, it is possible to facilitate the delivery of other substances to the deeper layers of the skin. These short peptides have small molecular weights, making them more likely to penetrate the skin. Among them, cell-penetrating peptides (CPPs) and skin-penetrating peptides (SPPs) are recognized as valuable delivery tools due to their high transduction efficiency and low cytotoxicity~\cite{copolovici2014cell,guidotti2017cell,gennari2016skin,hsu2011delivery}. 
A notable example is the TAT peptide (Trans-Activator of Transcription), a short peptide sequence (amino acids 47-57) that has been extensively studied and applied in cell penetration and drug delivery. When combined with the tripeptide GKH, the resulting TAT-GKH complex significantly enhances GKH’s skin penetration efficiency and is utilized in lipolytic cosmetic products~\cite{lim2003penetration}. 
Additionally, Chen et al. reported the use of the skin-penetrating peptide SPACE to enhance the topical delivery of high molecular weight hyaluronic acid (HA, molecular weight: 200-325 kDa)~\cite{chen2014topical}.
Carmichael and colleagues demonstrated that botulinum toxin type A (BoNT-A) could be administered subcutaneously or topically with a novel transdermal delivery peptide to reduce inflammation by inhibiting the activation of nociceptors in the skin~\cite{carmichael2010peptide}. Similar to small molecules, modifying the side chains of these peptides can also enhance their permeability. 
For example, the carrier peptide GHK can bind to Cu(II) and Mn ions in the skin~\cite{hur2020effect,hussain2007topical,errante2020cosmeceutical}. Through chemical modification, GHK can be linked to a palmitoyl group (a long-chain fatty acid), increasing its hydrophobicity and enhancing its interaction with the lipid bilayer, thereby improving its ability to penetrate the skin~\cite{arul2005biotinylated,jeong2019anti}.

Besides, liposomes are utilized to encapsulate peptides and proteins, enhancing their ability to penetrate the skin barrier more effectively~\cite{el2008liposomes,peralta2018liposomes}. 
Liposomes offer several advantages for the transdermal delivery of peptides and proteins, including improved permeability, protection of active ingredients, and controlled release. With the advent of novel liposome technologies, such as elastic liposomes, cationic liposomes, and pH-sensitive liposomes, delivery efficiency has been further enhanced~\cite{rai2017transfersomes,bouwstra2002skin,liu2013formati}

\section{Frontiers in molecular discovery for beauty design}
In \S\ref{sec:axes}, we have outlined the general guidelines of beauty design.
In this section, we focus on two directions particularly popular in beauty design---collagen formation and microhyaluronic acid application.

\subsection{The odyssey towards collagen}

Collagen, the most common type of protein found in skin as well as tendons and bones, assists fibroblasts forming in the dermis, promoting the growth of new cells and replacing dead skin cells.
Oral supplement of collagen has been linked to improved firmness, suppleness, moisture, smoothness, and elasticity of skin~\cite{pu2023effects, inoue2016ingestion, sugihara2015clinical}, although larger-scale study is needed to determine whether commercially available products are useful for long-term use.
Used topically, certain mixtures of amino acids, such as arginine and ornithine, have been proven to promote the disposition of collagen in wounds~\cite{williams2002effect, arribas2021effect, albaugh2017proline}.
Topically, the application of amino acids have also shown to be useful to boost collagen formation and skin moisture~\cite{trookman2009immediate}.

\subsection{Micro hyaluronic acids: potency and risks}
Micro hyaluronic acids with low molecular weight are well known for their ability to penetrate the skin effectively. These tiny molecules can go through the skin’s protective barrier and provide moisture directly to the deeper layers of the skin improving its elasticity and minimizing noticeable aging signs. This property is crucial in cosmetic dermatology, where hydration and skin suppleness are key to a youthful appearance. The effectiveness of micro hyaluronic acids in promoting skin hydration and reducing wrinkles has been substantiated by various studies, which highlight its efficacy as a humectant, with the ability to hold water up to a thousand times their weight and facilitate better absorption compared to their higher molecular counterparts~\cite{Bravo2022,polym14224833}. Advancements in skincare technology have leveraged the properties of micro hyaluronic acids to develop sophisticated delivery systems that maximize skin benefits. Techniques such as encapsulation in liposomes or binding to nanoparticles have been explored to further enhance the penetration and longevity of hyaluronic acids in skin layers. These innovations not only improve the immediate hydrating effects but also contribute to the long-term anti-aging benefits by maintaining higher levels of skin hydration over extended periods~\cite{polym14224833}.

Despite their benefits, microhyaluronic acids come with potential risks that must be managed carefully. Topical applications are generally safe but can occasionally cause skin irritation or mild allergic reactions in sensitive individuals. More invasive applications, such as injections, carry greater risks including infection, tissue necrosis, or adverse reactions at the injection site. It is vital that these treatments are administered by qualified healthcare professionals to minimize risks and ensure safety. Additionally, \citet{Bravo2022} conducted a comprehensive review and noted that extremely low molecular weights of hyaluronic acid may trigger inflammatory responses in the skin. It is necessary to have precise control over the molecular size used in skin care formulations.

Because of the growing usage of hyaluronic acid in the field of aesthetics, notably in injectable skin rejuvenation treatments, it has become essential to build a solid regulatory framework in order to protect the efficacy and safety of this substance. Regulatory scrutiny ensures that hyaluronic acid products, classified as medical devices, meet stringent safety standards and are marketed accurately regarding their benefits and potential side effects. The framework also highlights the importance of the administration by licensed practitioners and the need for ongoing research to support the claims made by manufacturers regarding the cellular effects of these injections~\cite{cosmetics11020054}. Hyaluronic acid usage regulations are made to protect the general public's health as well as satisfy the increasing market for anti-aging goods. This involves guaranteeing the chemicals' moral provenance and scientific backing. Strict regulatory measures are required as demand for these therapies rises in order to guard against abuse and guarantee that the goods live up to expectations without sacrificing safety.

\section{\textit{In silico} beauty design: recent advances.}

\paragraph{Efforts using machine learning to design peptides for skin care.}
Current trends in cosmetic molecule development closely align with those in medicinal chemistry~\cite{david2015pharmaceutical,pillaiyar2017skin}. 
In recent years, machine learning has played a significant role in peptide design and optimization, greatly improving the development efficiency of cosmetic peptides. 
Specifically, \citet{yue2024machine} reported using machine learning to predict the antimicrobial activity and hemolytic properties of peptides, leading to the design of a novel $\beta$-defensin-based antimicrobial peptide, IK-16-1. Additionally,  Hsueh et al. design a multifunctional peptides which can efficiently penetrate the cells and combined with melanin~\cite{hsueh2023machine}. In the area of skin penetration, De Oliveira et al. reported the development of BChemRF-CPPred, which extracts descriptors from peptide structures and sequences to distinguish cell-penetrating peptides (CPPs) from non-CPPs~\cite{de2021predicting}. The design of short peptides has become a prominent strategy in molecular discovery, primarily due to their strong targeting capabilities, high efficacy, stability, and low allergenicity~\cite{apostolopoulos2021global}. Although substantial progress has been made utilizing scientists’ domain knowledge, the \textit{de novo} design of short peptides remains relatively limited~\cite{aruan2023double,ganceviciene2012skin}.

\paragraph{Natural molecular discovery for skin care.}
Natural molecular discovery is emerging as a new trend in molecular design. This shift is largely driven by the perception that natural molecules are inherently safer and exhibit lower toxicity, aligning with consumer preferences and regulatory standards for safer cosmetic and therapeutic products. Consequently, most animal and plant extracts are widely recognized as safe, leading to the screening of many cosmetic molecules using chemical libraries derived from natural sources~\cite{martins2014marketed}. For example,  Bournez et al. employed homology modeling and molecular docking to screen 4,534 natural products for potential SIK2 inhibitors, aiming to enhance melanin production and achieve tanning effects without UV-induced damage~\cite{bournez2024virtual}. Besides, Su et al. compiled a database of 5,616 natural compounds and, through XP docking and molecular dynamics simulations, identified five compounds with promising dynamic binding properties. Subsequent in vivo experiments on zebrafish revealed that two components, Sennoside B (SB) and Sennoside C (SC), exhibited inhibitory effects on melanin production~\cite{su2024rapid}. 

\paragraph{Structural modification for cosmetic molecules.}
Structure modification is a primary method for developing new cosmetic molecules~\cite{ai2014novel}. Typically, chemical structures are altered to enhance their biological activity, stability, and Absorption, Distribution, Metabolism, Excretion, and Toxicity (ADMET) properties. In addition to these direct approaches, derivatives can also be generated based on known targets and established cosmetic molecules. Specifically, a typical example is the molecular modification of vitamins, to address the inherent instability and challenges with the topical application of vitamin C, formulation chemists have engineered derivatives such as ascorbyl 2-phosphate (SAP and MAP) and ascorbyl 6-palmitate (AA-Pal)~\cite{stamford2012stability}. These derivatives stabilize vitamin C within cosmetic formulations and enhance skin penetration by maintaining stability at a neutral pH and protecting the molecule from oxidation, facilitating their inclusion in various skincare products. Similarly, many researchers have modified natural products to enhance their bioactivity. Kumari et al. demonstrated that isoliquiritigenin and its derivatives, such as 5'-formylisoliquiritigenin and isoliquiritigenin dimers, exhibit higher binding affinity and more potent tyrosinase inhibition compared to isoliquiritigenin alone~\cite{kumari2023silico}. Molecular docking studies revealed that these derivatives have lower binding energies with tyrosinase, indicating their potential as more effective tyrosinase inhibitors for the prevention and treatment of melanogenesis and related hyperpigmentation disorders.

In addition to these methods, cosmetic formulations following molecular editing are further evaluated through their physical properties. Our article focuses on the use of chemical computational tools for prediction and combination. Therefore, in the subsequent Physical Toolbox section, we briefly introduce the relevant prediction tools, which will not be reiterated here.


\section{The physical toolbox: molecular dynamics (MD) simulation.}
Characterizing Equation~\ref{eq:boltzmann} by drawing samples while integrating Newton's second motion of law, MD simulation finds use in all schools of molecular modeling.
There have been a wide variety of mature and widely used pipelines for modeling the energy landscape of biomolecular systems of therapeutical interest, which can be seamlessly adopted to model those of cosmeceutical significance.

\paragraph{Modeling of cosmeceutical molecule-target binding.}
To design effective cosmetic molecules, molecular dynamics (MD) simulations~\cite{case2021amber,lee2016charmm,abraham2015gromacs, eastman2023openmm} can be employed to model the interactions between cosmeceutical molecules and their biological targets. To be more specific, MD simulations enable the accurate prediction of binding free energy, a key metric in assessing the strength and stability of the interaction between a small molecule, such as a cosmetic agent, and a target protein~\cite{mobley2017predicting}. One of these methods is commonly categorized as endpoint techniques, such as molecular mechanics generalized Born surface area (MM-GBSA) and molecular mechanics Poisson-Boltzmann surface area (MM-PBSA), to estimate free energy changes by evaluating the energy differences between the bound and unbound states with data sampled from the same molecular dynamics trajectory~\cite{wang2019end,genheden2015mm}.  In both approaches, the binding free energy is divided into several components, including enthalpic and entropic contributions. The former component is calculated as the sum of the gas-phase interaction energy between the ligand and protein and the solvation free energy. At the same time, the latter is computed as conformational entropy change related to ligand binding~\cite{doi:10.1080}. To enhance computational efficiency, these endpoint methods employ an implicit continuum solvent model, which significantly reduces the complexity of the calculations by omitting explicit water molecules from the MD simulations. While this approach accelerates calculations, it may also compromise the accuracy, as it overlooks the potentially significant role of specific water-mediated interactions.  
Besides, a key limitation of MM-PBSA and MM-GBSA is their treatment of entropy. These methods often neglect entropic contributions or handle them in a less rigorous manner. For instance, the configurational entropy is either approximated using simplified models or ignored altogether~\cite{wang2016calculating}. Therefore, MM-GBSA/PBSA exhibit moderate accuracy while being computationally efficient, which makes them well-suited for the rapid screening of potential small molecules in the early stages of drug discovery, such as anti-tumor and anti-virus drugs~\cite{hu2019rescoring,forouzesh2021effective,shridhar2021insilico}. For the calculation of complex interactions and highly flexible ligands, we developed the alanine scanning with generalized Born and interaction entropy (ASGBIE) method~\cite{liu2018computational,yan2017interaction}. Compared to MM-GBSA, ASGBIE provides more accurate binding free energy estimations derived from improved solvation energy calculations and a more rigorous consideration of entropic contributions, which are often neglected in MM-GBSA. Additionally, the method employs variable dielectric coefficients ranging from 1 to 10 for different categories of amino acids, including non-polar, polar, and charged groups, allowing for more nuanced electrostatic calculations. In a comparison of ASGBIE and MM-GBSA on 24 complex systems with 433 pairs of mutation, ASGBIE achieved a mean unsigned error (MUE) of 0.2 kcal/mol, outperformed the MUE of 1.0 kcal/mol for MM-GBSA~\cite{yan2017interaction}. Additionally, ASGBIE has been applied to predict hotspots with the highest contributions in binding free energy on a protein-peptide system, aiding peptide modifications to enhance binding affinity, and estimating relative binding free energy changes between wild-type and mutated peptides with reported accuracy~\cite{zhou2024efficient}.

More sophisticated methods can also be considered for estimating binding free energies. One important category is alchemical methods, including thermodynamic integration (TI)~\cite{kirkwood1935statistical} and free energy perturbation (FEP)~\cite{zwanzig1954high}. These methods estimate free energy changes by sampling intermediate states between initial and final configurations based on sampling configurations guided by the potential energy of the initial state. Two more extended functions were then developed based on the FEP equation. Bennett derived a Bennett acceptance ratio (BAR)~\cite{bennett1976efficient} function requiring sampling configuration at both states to minimize the expected square error of the free energy difference. Meanwhile, a multistate Bennett acceptance ratio (MBAR)~\cite{shirts2008statistically} function takes into account simulation data from multiple states, involving a more comprehensive sampling. These methods involve the calculation of free energy difference between two states by introducing a coupling parameter ($\lambda$) that varies from 0 to 1 to sample configurations for both states, either based on the ensemble average of the state 0 or that of both states, creating a series of intermediate states that improve phase space overlap and facilitate a smoother transformation. The utilization of soft-core potential further avoids the dispersed energy distribution on endpoints of 0 and 1. Furthermore, alchemical methods have been extensively applied in estimating binding free energy between target proteins and small molecules, such as large-scale screenings or docking poses evaluation~\cite{bhati2022large,goldfeld2015docking,kaus2015deal}. Some approaches combine alchemical methods with replica exchange solute tempering (FEP/REST), further improving accuracy by overcoming energy barriers associated with local minima~\cite{bourion2021psychological,lockhart2023can,ngo2020oversampling}. Several software platforms support alchemical methods. For instance, AMBER18 enables GPU-accelerated TI calculations, and AMBER20 package was evaluated for its accuracy on a dataset of around 200 ligand mutations spanning 8 protein targets, achieving a mean unsigned error (MUE) of 0.84 kcal/mol~\cite{lee2020alchemical}. Similarly, a commercial software FEP+ exploited FEP/REST approach on the same dataset and achieved a comparable MUE of 0.86 kcal/mol~\cite{wang2015accurate}. These alchemical approaches offer more efficient sampling of the protein-ligand phase space and have demonstrated strong correlations with experimental data, making them reliable for accurately predicting the binding free energy of small cosmeceutical molecules interacting with their targets. Another category of methods includes enhanced sampling techniques like umbrella sampling and metadynamics, which are useful for large-scale molecular systems~\cite{williams2018free}. However, these methods significantly increase computational demand due to the exponential growth in the number of molecules and interactions, necessitating the use of high-performance computing (HPC) resources and substantial storage capacity.

\paragraph{Physical property prediction.}
Usually, Cosmetic compounds necessitate a thorough evaluation of ADMET profiles to guarantee their suitability for topical application~\cite{martins2014marketed}. This comprehensive assessment ensures that active ingredients are effectively absorbed through the skin barrier, distributed appropriately within the skin layers, metabolized without generating harmful by-products, and excreted without accumulation that could lead to adverse effects. Traditionally, ADMET evaluation has relied primarily on experimental methods to determine these key pharmacokinetic and toxicological properties of drugs or active molecules. Recently, MD simulations have emerged as an important tool for aiding in the computational analysis of ADMET~\cite{srivastava2025synthesis,balkrishna2024silico}. By simulating the behavior of molecules within cell membranes and skin tissues, researchers can predict how these molecules penetrate the skin barrier, their distribution within the skin, and their interactions with proteins. Additionally, MD simulations can be combined with other computational methods, such as machine learning and computational fluid dynamics, to predict the fluid properties of products under various conditions, including viscosity, surface tension, and diffusion coefficients. These predictions can effectively guide the design of cosmetic formulations, reduce development costs, and improve the application performance and stability of products.

\section{The machine learning toolbox}
\paragraph{Graph-based machine learning.}
Graph neural networks~\cite{DBLP:journals/corr/KipfW16, xu2018powerful, gilmer2017neural, hamilton2017inductive, battaglia2018relational, wang2024nonconvolutionalgraphneuralnetworks} have be come the \textit{de facto} workhorse of all schools of molecular modeling.
A graph neural network can be most generally defined as one adopting a layer-wise updating scheme that aggregates representations from a node's neighborhood and updates its embedding:
\begin{equation}
\label{eq:master-local}
\mathbf{X}'_v = \phi (\mathbf{X}_v, \rho (\mathbf{X}_u, \hat{A}_{uv}, u \in \mathcal{N}(v) )),
\end{equation}
where $\phi, \rho$ are the \textit{update} and \textit{aggregate} function, repsectively.
Omitting the nonlinear transformation step $\phi$ ubiquitous in all neural models, and assuming a convolutional \textit{aggregate} function, $\rho = \operatorname{SUM}$ or $\rho = \operatorname{MEAN}$, a graph neural network layer is characterized by the aggregation/convolution operation that pools representations from neighboring nodes, forming an intermediary representation $\mathbf{X}'$, which on a global level, with activation function $\sigma$ and weights $W$, can be written as: $\mathbf{X}' = \sigma(\hat{A}\mathbf{X}W)$.
The primary difference among architectures amounts to distinct effective adjacency matrix $\hat{A}$.
The most classical examples include: graph convolutional networks (GCN)~\cite{DBLP:journals/corr/KipfW16}, which normalizes $\hat{A}$ by the in-degree of nodes $D_{ii} = \sum_j A_{ij} $ and
graph attention networks (GAT)~\cite{velivckovic2017graph}, which takes $\hat{A}$ as the attention score;
\begin{equation}
\label{eq:gcn}
    \hat{A}_\mathtt{GCN} = D^{-\frac{1}{2}} A D^{-\frac{1}{2}};
    \hat{A}_{\mathtt{GAT}, ij} = \operatorname{Softmax}(
        \sigma(
            \operatorname{NN}(
                \mathbf{x}_i ||
                \mathbf{x}_j
            )
        )
    );
\end{equation}
The resulting representation is rich and naturally permutation-invariant, providing a strong starting point for downstream tasks.

\paragraph{Generative modeling.}
Once we have a reliable model for the prediction of activities given the structure of the graph, we can couple it with a generative model for the true \textit{de novo} design of molecules and peptides.
This can be achieved through a principled, probabilistic manner, where the conditional distribution in Equation~\ref{eq:posterior-predictive} is replaced by the distribution of the chemical graphs without measurements:
\begin{equation}
p(\mathcal{G} | \mathcal{M}) = \int \operatorname{d} \Theta p(\mathcal{G}, \Theta | \mathcal{M}) p(\Theta | \mathcal{M}),
\end{equation}
and the model is typically trained by maximizing the (lower bound of) the likelihood of data $p(\mathcal{G} | \mathcal{M})$.
For the generation of molecular structures, common approaches include variational autoencoder (VAE)-based models, where molecules are generated as strings~\cite{gomez2018automatic}, graphs~\cite{kipf2016variationalgraphautoencoders}, or a Markov-decision process to construct from fragments~\cite{jin2019junctiontreevariationalautoencoder};
flow-based models~\cite{Zang_2020};
and diffusion-based models~\cite{liu2023generativediffusionmodelsgraphs}.
The conformations of molecules can also be simultaneously or conditionally generated in the 3-dimensional space~\cite{hoogeboom2022equivariantdiffusionmoleculegeneration}.

\paragraph{Active learning.}
With some over-simplification, molecular discovery can be abstracted as a sequential active learning process.
Given a finite pool of candidate chemical compounds, represented by chemical graphs $\mathcal{F} = \{ \mathcal{G}_i, i=1,2,\dots N\}$, each associated with a \textit{potency} $y(\mathcal{G}_i) \in \mathbb{R}$, of which a noisy observation is expensive to acquire, we are interested in finding the compound with the highest potency $\mathcal{G}^* = \operatorname{argmax}_iy(\mathcal{G}_i)$ with fewest function evaluations.
This can be achieved via \textit{active learning}, starting with an empty portfolio $\mathcal{P} = \emptyset$, and in each round, choosing a compound from the candidate pool, $\mathcal{F}$, based on a decision informed by the current portfolio, $\hat{\mathcal{G}} = D(\mathcal{F} | \mathcal{P}), \hat{\mathcal{G}} \in \mathcal{F}$.
Subsequently, we subtract this compound from the candidate pool to add it to the portfolio, $\mathcal{F} \leftarrow \mathcal{F} \setminus \{\hat{\mathcal{G}}\}, \mathcal{P} \leftarrow \mathcal{P} \cup \{\hat{\mathcal{G}}\}$.

Bayesian optimization (BO) is a powerful approach in active learning where the decision $D(\mathcal{F}|\mathcal{P})$ is constructed in a modular way.
Firstly, we extract fixed $D$-dimensional features from the compounds to form a \textit{representation}: $H=h(\mathcal{G}) \in \mathbb{R}^D$;
secondly, this representation is used to construct a \textit{predictive posterior distribution} (Eqaution~\ref{eq:posterior-predictive}) with \textit{uncertainty quantification}: $p(y|\mathcal{G}; \Theta), \Theta = \theta(H)$, whose parameters $\Theta$ is is mapped from the graph representations.
Lastly, an \textit{acquisition} function, $\alpha$, is applied on this predictive posterior to form a score, on the basis of which a decision is made: $\hat{G} = D(\mathcal{F} | \mathcal{P}) = \operatorname{argmax}_{\mathcal{G} \in \mathcal{F}}\alpha(p(y|\mathcal{G}))$.
This process can be executed iteratively until a satisfactory compound has been identified.

\paragraph{Large-language models.}
Large language models (LLMs)---neural networks composed primarily of transformer~\cite{vaswani2023attentionneed}---have shown surprising promise~\cite{touvron2023llamaopenefficientfoundation, openai2023chatgpt, brown2020languagemodelsfewshotlearners, bender2021dangers, ni2024harnessingearningsreportsstock} in representing and generating text-structured data and gained popularity very quickly in recent years.
There have been a plethora of LLMs for chemical tasks~\cite{jablonka2023gpt}, the most popular ones include \citet{yu2024llasmoladvancinglargelanguage, zhang2024chemllmchemicallargelanguage}, which have been fine-tuned on instruction datasets~\cite{zhang2024instructiontuninglargelanguage}.
LLMs can be integrated into various workflows discussed here, from planning experiments to be carried out in an active learning setting, to writing code to prepare biomolecular systems for simulations.

\section{Roadmap: Potential use cases}
\label{sec:roadmap}
Having reviewed the goals and challenges in the beauty design industry and the powerful methodologies in our physical and statistical toolboxes, in this section, we propose three projects that are immediately feasible, to showcase the potential power of combining MD and ML approaches in beauty design.

\paragraph{The molecular modification of arbutin via multiobjective active learning.}
Inhibiting the activity of tyrosinase, an essential enzyme for melanin synthesis in skin, arbutin has been ubiquitously used as skin-lightening agents targeting hyperpigmentation issues.
Usually extracted from the leaves of plants, such as the bearberry, the efficacy of this natural product is bottlenecked by its inability to penetrate the skin to reach the melanocytes~\cite{masyita2021molecular}.
We propose to develop arbutin derivatives displaying similar efficacy while showing significantly greater lipophilicity for better skin penetration.

First, docking and molecular dynamics simulations will be carried out to better understand the binding mechanism of arbutin.
With these structural insights, we will carry out molecular modification, for example using fragment-based tools such as \citet{zhang2024fraggrow}, to replace functional groups on arbutin.
This can be done in an ultra-high throughput manner, generating a large library of arbutin-derivatives.
Multitask Bayesian optimization algorithms can then design experiments to be carried out sequentially, to optimize for both lipophilicity and melanocytes-inhibiting activity, which can be accessed either experimentally or using free energy calculations.

\paragraph{Optimizing the ADMET profile of hyluronic acid.}
The extremely small molecular weight of hyaluronic acid is both a blessing and a curse---it enables the easy penetration of the skin while also triggering inflammatory reactions~\cite{Bravo2022}.
The chemical modification of hyaluronic acid~\cite{bokatyi2024chemical} has been proven to be an effective avenue towards efficient drug delivery.
Using the modification strategies outlined in \citet{bokatyi2024chemical}, we can chemically decorate the backbone with inactive functional groups to moderately increase the molecular weight of hyaluronic acid, while not altering its ability to hold water.
These modifications can be proposed either by generative models or fragment-based structure enumeration and library generation tools such as \citet{polishchuk2020crem}, while the physical properties and the ADMET profile can be predicted using a machine learning-enabled QSAR model.

\paragraph{Cosmetic peptide design. }
Peptides are currently predominantly composed of oligopeptides (containing fewer than 10 amino acids), forming a vast polypeptide space. The design of synthetic peptides still relies heavily on traditional, resource-intensive methods that are both time-consuming and laborious. These conventional approaches often result in peptides that lack the necessary modifications to enhance their skin penetration and bioactivity. However, advancements in technology have made peptide sequence prediction and optimization more efficient and precise. In terms of de novo peptide design, researchers can now leverage diffusion models trained on medicinal peptide datasets to capture the complex patterns of peptide sequence distribution and functional properties, enabling the generation of novel peptides with desired characteristics. Additionally, protein function prediction tools, such as DeepFRI, can be employed to predict the corresponding protein functions of these peptides~\cite{gligorijevic2021structure}. From a peptide modification perspective, targeted chemical modifications can be introduced to enhance the permeability and stability of peptides. For instance, acetyl tetrapeptide-15, a chemically modified version of Tyr-Pro-Phe-Phe, demonstrates how modifications like cyclization, lipid chain conjugation, or terminal glycosylation can be applied. These modified peptides can then be evaluated using molecular dynamics (MD) simulations. Through iterative cycles of model evaluation and screening, the efficacy and safety of the generated peptides can be validated.

\paragraph{Towards precision beauty design: personalized mixture for cosmetics and skin care.}
Drawing inspirations from the rich literature of \textit{drug combinations}~\cite{jaaks2022effective}, which characterizes the synergistic effect among multiple small molecule drugs on multi-factorial diseases, we propose to design a system to design formulae of skin-care and cosmetic mixtures that cater to personal conditions and requirements.
For instance, such a system can take a picture of a person's face, and propose various solutions for lipstick mixtures, which contain waxes, oils, pigments, moisturizers, and antioxidants, and simulate their effects based upon techniques such as spectra prediction.



\section{Conclusions}
From this perspective, we discuss the contributions of computational tools, including molecular dynamics and machine learning, to the discovery and development of cosmetic molecules. Cosmetic molecules exhibit a high degree of similarity to drug molecules, and the development of corresponding cosmetic molecules is still in its early stages, offering significant potential for further growth. At the end of this paper, we present a roadmap outlining several feasible approaches.
These approaches represent the future of small molecules, proteins, and physics-based comprehensive product development, respectively.

\section*{Disclosures. }
YW has limited financial interests in Flagship Pioneering, Inc. and its subsidiaries.

\section*{Funding. }
YW acknowledges support from the Schmidt Science Fellowship, in partnership with the Rhodes Trust, and the Simons Center for Computational Physical Chemistry at New York University.

\bibliography{main}
\end{document}